\documentclass[conference]{IEEEtran}
\IEEEoverridecommandlockouts
\usepackage{cite}
\usepackage{amsmath, amssymb, amsfonts, mathrsfs}
\usepackage{algorithmic}
\usepackage{times}
\usepackage[margin=1in]{geometry}
\usepackage[normalem]{ulem} 
\usepackage{graphicx, color}
\usepackage{textcomp}
\usepackage{xcolor}
\usepackage{multirow}

\newcommand{\eat}[1]{}

\begin{document}

\title{\huge{Solving Newton's Equations of Motion with Large Timesteps using Recurrent Neural Networks based Operators}\\}

\author{\IEEEauthorblockN{
JCS Kadupitiya,
Geoffrey C. Fox,
Vikram Jadhao
}
\IEEEauthorblockA{\textit{Intelligent Systems Engineering} \\
\textit{Indiana University}\\
Bloomington, Indiana 47408 \\
\{kadu,gcf,vjadhao\}@iu.edu}
}

\maketitle

\begin{abstract}
Classical molecular dynamics simulations are based on solving Newton's equations of motion. Using a small timestep, numerical integrators such as Verlet generate trajectories of particles as solutions to Newton's equations. We introduce operators derived using recurrent neural networks that accurately solve Newton's equations utilizing sequences of past trajectory data, and produce energy-conserving dynamics of particles using timesteps up to 4000 times larger compared to the Verlet timestep. We demonstrate significant speedup in many example problems including 3D systems of up to 16 particles.
\end{abstract}



\section{Introduction}

Newton's equations of motion \cite{newton} are the basis of powerful computational methods such as classical molecular dynamics (MD) that are used to understand the microscopic origins of a wide range of material and biological phenomena 
\cite{alder1959studies,frenkel}.
In the MD method, Newton's equations are integrated for a system of many particles using numerical integrators such as Verlet \cite{verlet1967computer} to produce particle trajectories. The time evolution is performed one small timestep at a time for long times to accurately sample enough representative configurations in order to extract useful information. 
Consider the 2nd order ordinary Verlet integrator $\vec{x}(t + \Delta) = 2\vec{x}(t) - \vec{x}(t-\Delta) + \Delta^2 \vec{f}(t) / m$ that updates the current position $\vec{x}(t)$ of a particle of mass $m$ at time $t$ to position $\vec{x}(t + \Delta)$ after timestep $\Delta$ using the previous position $\vec{x}(t - \Delta)$ and the force $\vec{f}(t)$ at time $t$. 
This integrator produces an error of $O(\Delta^4)$ in each local update and incurs a global error of $O(\Delta^2)$ \cite{andersen1983rattle,frenkel}.
These errors are reduced by choosing a small $\Delta$ which often makes the simulations computationally expensive.

The ordinary Verlet integrator requires a sequence of 2 positions ($\vec{x}_{t-\Delta},\vec{x}_t$) to update the particle position using other quantities (e.g.,  $\vec{f}$ and $m$). These quantities can be inferred using the information encoded in a long sequence of positions such that the time evolution can be done with only the history of positions as input.
We illustrate this with a 1-dimensional example of a particle experiencing simple harmonic motion governed by the force $f = -k x$. One can show that the particle position can be evolved to $t+\Delta$ using a sequence of 3 positions via the function $\mathcal{V} = x_{t-\Delta}^{-1}\left( x_{t}^2 - x_{t-\Delta}^2 + x_{t}x_{t-2\Delta} \right)$, which also incurs a global error of $O(\Delta^2)$.
This idea generalizes for higher-order integrators \cite{butcher2016numerical} and many-particle systems such that the time evolution can be performed via $\mathcal{V}\left(\vec{x}_t, \vec{x}_{t-\Delta}, \ldots \vec{x}_{t-s\Delta}\right)$ that takes a sequence of $s$ positions.
The longer history of input positions enables integrators to perform accurate time evolution with a larger $\Delta$, however, generally at the expense of higher computing costs per timestep.

The use of deep learning in sequence processing and time series prediction problems has been well studied by the industry for different applications including voice recognition and translation \cite{wu2016google}, pattern recognition in stock market data \cite{chong2017deep}, and ride-hailing  \cite{gcfref6}. Recurrent neural networks (RNNs) are established deep learning tools in these applications. 
In this work, we develop RNN based operators to perform accurate time evolution of one-particle and few-particle systems utilizing sequences of past trajectories of particles. 
The RNN-based operators are trained using the ground truth results obtained with the Verlet integrator. They possess a complex mathematical structure described with up to $100,000$ parameters.
We demonstrate that the network complexity enables the operators to perform time evolution of systems of up to 16 particles for a wide range of force fields using timesteps that are up to $4000\times$ larger than the baseline Verlet timestep. The relatively small time for inferring the positions as predictions of the deep learning model keeps overhead costs low and we demonstrate significant net speedups for larger timesteps.

\begin{figure*}[t]
\centering
\includegraphics[width=0.98\textwidth]{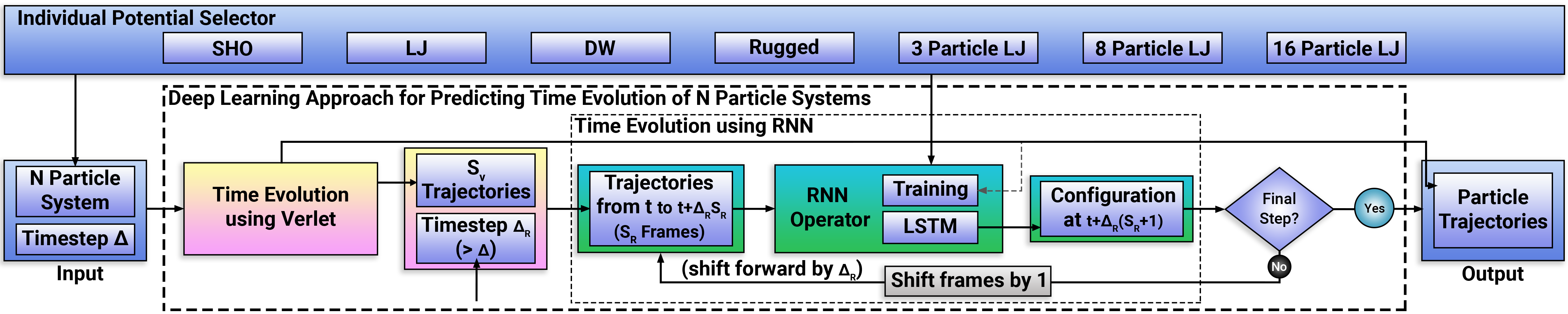}
\caption{Overview of the deep learning approach involving recurrent neural networks (RNN) based operators to solve Newton's equations of motion and predict dynamics of $N$ particles.
}
\label{fig:RNN_overview}
\end{figure*}

Machine learning has been used recently to enhance the performance of MD simulations  \cite{ferguson2017machine,butler2018machine,gcfref1,Sharp10943,glotzer2017,guo2018adaptive,botu2015adaptive,kadupitiya2020ml,long2015machine,moradzadeh2019molecular,sun2019deep,aspuru2019,kadupitiya2019machine,kadupitiya2020machine,wang2019machine}.  
Deep learning approaches that learn differential equations and replicate the outputs of numerical integrators \cite{raissi2018hidden, long2017pde, chen2018neural, endo2018multi, breen2019newton, chen2019symplectic, shen2017essential, long2017pde,raissi2019physics,bar2019learning,shen2020deep,raissi2018multistep} are of particular relevance to our work.
Recently, such efforts have focused on solving differential equations with discretization steps larger than the baseline \cite{shen2020deep,raissi2018multistep}.
Most of these approaches have been evaluated on relatively simple 1D systems. We also note that a deep learning approach has been recently proposed to determine the evolution of configurations described by a few collective variables characterizing the system dynamics \cite{tsai2020learning}.
On the other hand, we note the active work in the development of approaches that do not rely on deep learning, such as the multiple timestep methods, for simulating the dynamics of complex systems with large timesteps \cite{minary2004long,morrone2011efficient,leimkuhler2013stochastic,chen2018molecular}. 

\section{Recurrent Neural Network based Operators for Predicting Dynamics of Particles}

Figure \ref{fig:RNN_overview} shows the overview of our deep learning approach to learn the time evolution operator that evolves the dynamics of an $N$-particle system.
We begin by selecting the potential energy function governing the dynamics of the particles. In addition to the potential energy, the $N$-particle system is specified by particle masses and the initial positions and velocities of the particles.
The attributes of the $N$-particle system together with the Verlet timestep $\Delta$ form the input.
The input system attributes are fed to the Verlet integrator to simulate the dynamics with timestep $\Delta$ up to $S_{\mathrm{V}}$ computational steps.
Out of the full trajectory data (e.g., positions and velocities) up to $S_{\mathrm{V}}$ steps, $S_{\mathrm{R}}$ number of configurations (frames) separated by $\Delta_{\mathrm{R}}$ are distilled. Note that this requirement to kickstart the time evolution using the Verlet integrated enforces $S_{\mathrm{V}} = \Delta_{\mathrm{R}}(S_{\mathrm{R}} - 1)/\Delta$.
Using this initial sequence of particle configurations of length $S_{\mathrm{R}}$, a trained recurrent neural network based operator $\mathscr{R}$ predicts the time evolution of the system after timestep $\Delta_{\mathrm{R}}$. Then, the input sequence to $\mathscr{R}$ is backward (left) shifted to discard the oldest time frame, and the latest frame predicted by $\mathscr{R}$ is appended to the front end (right) of the sequence. The adjusted input sequence is fed back to $\mathscr{R}$ to evolve the system $\Delta_{\mathrm{R}}$ further in time and the same process is repeated until the end of the simulation. 
The time evolution results in the output comprising the trajectories of the particles.

As shown in Figure \ref{fig:RNN_overview}, $\mathscr{R}$ is trained using the ground-truth particle trajectories generated via the velocity Verlet integrator with small timestep $\Delta =0.001$ for the system specified by the selected potential energy function. The velocity Verlet integrator updates the configuration of particles via two steps, which we describe for the case of a single particle; extension to many particles is straightforward. First, the position $\vec{x}(t)$ of a particle of mass $m$ at time $t$ is evolved a timestep $\Delta$ forward in time:
\begin{equation}\label{eq:posupdate}
\vec{x}(t + \Delta) = \vec{x}(t) + \Delta \vec{v}(t) + 0.5\Delta^2 \vec{f}(t) / m,
\end{equation} 
where $\vec{v}(t)$ and $\vec{f}(t)$ are the current velocity and force at time $t$ respectively. 
Next, the velocity $\vec{v}(t)$ at time $t$ is updated to $\vec{v}(t+\Delta)$:
\begin{equation}\label{eq:velupdate}
\vec{v}(t + \Delta) = \vec{v}(t) + 0.5\left(\Delta/m\right) \left(\vec{f}(t)+\vec{f}(t+\Delta)\right),  
\end{equation}
where $\vec{f}(t+\Delta)$ is the force computed at time $t+\Delta$ using the updated position of the particle evaluated in Equation \ref{eq:posupdate}. The time evolution moves forward following Equations \ref{eq:posupdate} and \ref{eq:velupdate} with $\vec{x}(t+\Delta)$ and $\vec{v}(t+\Delta)$ as current position and velocity respectively.

We emphasize that our approach trains separate recurrent neural networks for furnishing the time evolution of systems described by different functional forms of potential energy. For example, if a 1D simple harmonic potential $1/2kx^2$ is selected as the potential energy function, $\mathscr{R}$ learns to predict the dynamics of one particle in a harmonic potential for unseen values of $k$, however, it is not trained to predict the time evolution of a particle in a qualitatively different potential energy such as a double well potential.
The recurrent neural network based operators are at the heart of our approach. 
In order to understand how these operators are designed and trained, we first briefly describe the key characteristics of recurrent neural networks.


\subsection{Recurrent neural networks}
Recurrent neural networks (RNNs) process input sequence data and maintain a vector $\vec{h}_t$ known as the ``hidden state'' for each recurrent cell to model the temporal behavior of sequences through directed cyclic connections between cells. $\vec{h}_t$ is updated by applying a function $F$ to the previous hidden state ($\vec{h}_{t-1}$) and the current input ($\vec{x}_{t}$).
The cells are arranged in a fashion where they fire when the right sequence is fed. 
A common choice for $F$ is the Long Short Term Memory (LSTM) units \cite{hochreiter1997long}. 
There are several architectures of LSTM units. An often employed architecture consists of a cell (the memory part of the LSTM unit) and three ``regulators'', usually called gates, that regulate the flow of information inside the LSTM unit. An input gate ($i_t$) controls how much new information is added from the present input ($x_t$) and past hidden state ($h_{t-1}$) to the present cell state ($c_{t}$). A forget gate ($f_t$) decides what is removed or retained and carried forward to $c_{t}$ from the previous cell state ($c_{t-1}$). An output gate ($o_t$) decides what to output as the current hidden state ($h_t$) from the current cell state.
The LSTM formulation can be expressed as:
\begin{eqnarray} 
f_t &=\sigma_g (W_f x_t + U_f h_{t-1} + b_f) \nonumber \\ 
i_t &=\sigma_g (W_i x_t + U_i h_{t-1} + b_i) \nonumber \\
o_t &=\sigma_g (W_o x_t + U_o h_{t-1} + b_o) \nonumber\\
\tilde{c_t} &=\sigma_h (W_c x_t + U_c h_{t-1} + b_c) \nonumber \\
c_t &= f_t \circ c_{t-1} + i_t \circ \tilde{c_t} \nonumber \\
h_t &= o_t \circ \sigma_h(c_t).
\end{eqnarray}
Here, $x_t \in \mathbf{R}^d$ is the input vector to the LSTM unit, $f_t \in \mathbf{R}^h$ is the forget gate's activation vector, $i_t \in \mathbf{R}^h$ is the input gate's activation vector, $o_t \in \mathbf{R}^h$ is the output gate's activation vector, $h_t \in \mathbf{R}^h$ is the hidden state vector also known as the output vector of the LSTM unit, $c_t \in \mathbf{R}^h$ is the cell state vector, and $\circ$ is the Hadamard  product operator.  $W \in \mathbf{R}^{h\times d}$ and $U \in \mathbf{R}^{h\times h}$ are the weight matrices and $b \in \mathbf{R}^h$ are the bias vector parameters which need to be learned during training. $\sigma_g$ and $\sigma_h$ represent sigmoid function and hyperbolic tangent functions respectively. $d$ and $h$ refer to the number of input features and the number of hidden units respectively.

We now describe how RNNs with LSTMs can be used to design operators that process sequences of particle configurations to evolve the associated system forward in time. 

\begin{figure}[hbt]
\centering
\includegraphics[width=0.48\textwidth]{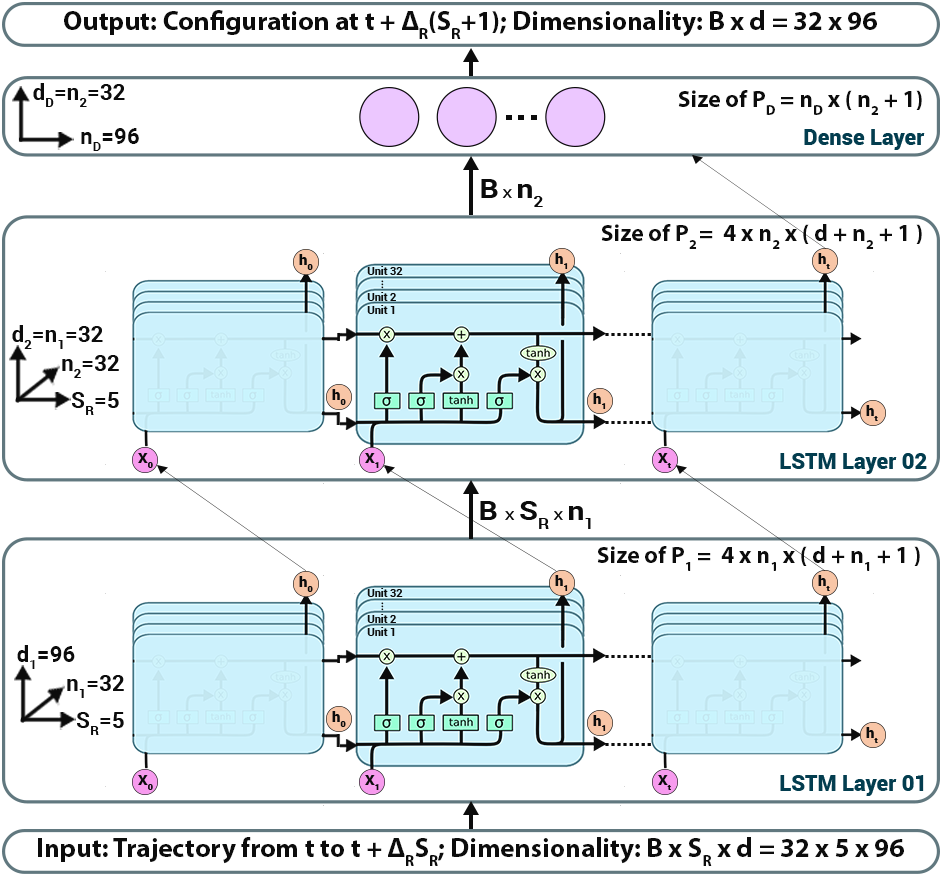}
\caption{RNN operator performing the time evolution of an $N$ particle system characterized by features of size $d$. The operator evolves the system one $\Delta_{\mathrm{R}}$ timestep forward in time using a sequence of length $S_R$ of past trajectory states. The update is shown for an input system of $N=16$ particles in 3D for which the feature size is $d=96$. Parameters associated with Long Short Term Memory (LSTM) units are also shown. These parameters are defined in the main text.}
\label{fig:LSTMArchitecture}
\end{figure}


\subsection{RNN-based Time Evolution Operators}

We design operators $\mathscr{R}$ using RNNs with LSTM units that process a sequence of past positions and velocities as input and generate the future positions and velocities of the particles.
Each component of the particle position and velocity vectors is identified as a feature. The feature size associated with the inputs and outputs for $N$ particles in $\mathcal{D}$ physical dimensions is $d = N \times \mathcal{D} \times 2$. For example, for $N=16$ particles interacting in 3D, $d = 96$. 
For a system specified by the selected potential energy function, operator $\mathscr{R}$ predicts the future position and velocity vectors of the particles at time $t+\Delta_{\mathrm{R}}$ by employing a sequence of length $S_{\mathrm{R}}$ of positions and velocities $\{x,v\} = \vec{x}_t, \vec{v}_t, \vec{x}_{t -\Delta_{\mathrm{R}}}, \vec{v}_{t -\Delta_{\mathrm{R}}}, \ldots, \vec{x}_{t - S_{\mathrm{R}}\Delta_{\mathrm{R}}}, \vec{v}_{t - S_{\mathrm{R}}\Delta_{\mathrm{R}}}$ up to time $t$. 
$\mathscr{R}$ is expressed as $\mathscr{R} [\{x,v\}] = \mathscr{D} [ \mathscr{L}_2 [ \mathscr{L}_1 [ \{x,v\} ] ] ]$, where $\mathscr{D}$, $\mathscr{L}_1$ , $\mathscr{L}_2$ are the operators associated with the dense layer, the first LSTM layer, and the second LSTM layer of the RNN respectively. 

The layers are stacked up on each other (Figure \ref{fig:LSTMArchitecture}) such that the output of one (e.g., $\mathscr{L}_1$) becomes the input for another ($\mathscr{L}_2$). 
Each LSTM layer consists of $n$ number of LSTM units and contains a set of parameters in the form of weights, biases, and activation functions. For example, $\mathscr{L}_1$ has $n_1$ LSTM units and is characterized with weights $W$ and $U$, and bias $b$. It takes input feature vector $\{x, v\}$ and outputs hidden state vectors $\{h\}$ which are fed as input to the $\mathscr{L}_2$ layer characterized with its own set of weights and biases. A similar connection is made between $\mathscr{L}_2$ and the dense layer $\mathscr{D}$.
Post training, these layers acquire optimal values for all the parameters, and the operator $\mathscr{R}$ emerges as:
 \begin{equation}  \label{eq:RNN_Operator2}
   (\vec{x}_{t+\Delta_{\mathrm{R}}}, \vec{v}_{t+\Delta_{\mathrm{R}}}) = \mathscr{D} \left[\mathscr{L}_2 \left[\mathscr{L}_1 \left[ \{x,v\}, \{P_1\}\right], \{P_2\}\right], \{P_D\}\right],
\end{equation}
where $\{P_1\}$, $\{P_2\}$, $\{P_D\}$ are optimized parameters associated with LSTM layer 1, LSTM layer 2, and the dense layer respectively. 
$\mathscr{R}$ has a complex mathematical structure characterized with up to $100,000$ parameters.

A similar process can be used to design operators that take a sequence of past positions as input and generate the future positions of the particles.

\subsection{Operator Training and Implementation Details}



We now discuss the details of the training and implementation of RNN-based operators. 
These operators are created in TensorFlow with LSTM layer 1, LSTM layer 2, and final dense layer of sizes (number of hidden units) $n_1$, $n_2$, and $n_D$ respectively.
While training an operator $\mathscr{R}$ for a specific potential energy function, a $B \times S_{\mathrm{R}} \times d$ dimensional vector comprising a sequence of positions and velocities $\{x,v\}$ is fed to an operator $\mathscr{R}$ as input. Here, $B$ is a training parameter denoting the batch size, $d$ is the feature size, and $S_{\mathrm{R}}$ is the aforementioned sequence length. During the testing phase, $B=1$.
All the parameters $\{P\}$ including the weights and biases describing the layers are optimized with an error backpropagation algorithm, implemented via stochastic gradient descent. Adam optimizer is used to optimize the error backpropagation. Outputs of the LSTM layers are wrapped with the $\tanh$ activation function. No activation function is used in the final dense layer. 
The mean square error (MSE) between target and predicted trajectories is used for error calculation. 

The parameters $\{P\}$ of the operator $\mathscr{R}$ are saved and loaded using Keras library \cite{chollet2015keras}.
Values of $n_1$, $n_2$, and $n_D$ are chosen depending on the problem complexity and data dimensions. 
For example, in the case of 16 particles interacting with LJ potential in 3D with periodic boundary conditions (feature size $d=96$), by performing a grid search of the parameters $\{P\}$ using Scikit-learn library \cite{buitinck2013api}, hyper-parameters such as the number of units for each of the two LSTM layers ($n_1$, $n_2$), number of units in the final dense layer ($n_D$), batch size ($B$), and the number of epochs are optimized to 32, 32, 96, 256, and 2500 respectively. 
The learning rate of Adam optimizer is set to 0.0005 and the dropout rate is set to 0.15 to prevent overfitting. Both learning and dropout rates are selected using a trial-and-error process. The weights in the hidden layers and in the output layer are initialized for better convergence using a Xavier normal distribution at the beginning \cite{glorot2010understanding}.  

Standard practices are followed to train the RNN-based operators to accurately predict trajectories while avoiding overfitting. First, the operator $\mathscr{R}$ is trained using all the training data. As expected, this model is generated in the overfitted region and it predicts results with small errors for the training samples but does not provide the same accuracy for the validation data. Next, we progressively constrain the model by reducing the number of parameters and introducing dropouts, until we obtain a similar level of low errors for samples in both training and validation datasets. Any signature of overfitting the RNN model would result in the trajectory predictions for systems in the training dataset with much lower errors compared to the errors for predictions associated with systems in the validation dataset. 

We experimentally find the minimum number of hyperparameters required to keep the RNN models well-generalized and avoid overfitting by finding the optimum point in the bias-variance risk curve for the training and testing error, and introducing dropout regularization between intermediate layers of the RNN while training. Large errors obtained during the prediction of trajectories in the validation and testing datasets also alert us to the case of insufficient training samples. In general, we find that the required number of training samples depends on the complexity of the potential energy landscape. For example, in the case of 1D systems, the number of training samples required to train operator $\mathscr{R}$ to predict the trajectory of a particle in the rugged potential is $1.3\times$ the training samples needed by the operator $\mathscr{R}$ designed to predict the dynamics of the same particle in a double well potential. Similarly, training the RNN operator to accurately predict the dynamics of the 3D many-particle systems required $10\times$ more training samples compared to the operators trained to predict dynamics for 1D systems. 

Prototype implementation of RNN-based operators written using Python/C++ is publicly available on GitHub \cite{github.RNNMD}.

\section{Results and Discussion}


One particle experiments in 1D are performed for 4 potential energy functions: simple harmonic, double well, Lennard-Jones, and rugged (see Appendix, Figure \ref{fig:potential_plots}). 
Experiments on $N-$particle systems with $N=3, 8, 16$ particles in 3D are performed on particles interacting with Lennard-Jones potentials in a cubic simulation box with periodic boundary conditions.
We adopt units such that the input parameters and predicted quantities are around 1.

In all experiments, RNN-based operators trained with a sequence length of $S_{\mathrm{R}} = 5$ are used to  perform the time evolution. Operators trained with smaller $S_{\mathrm{R}} =3$ or $4$ are only able to accurately propagate the dynamics for timestep $\Delta_{\mathrm{R}} \lesssim 10\Delta$, where $\Delta$ is the baseline Verlet timestep (see Appendix, Figure \ref{fig:sequancelength}). 
Trained with $S_{\mathrm{R}} = 5$, operators produce accurate dynamics for timesteps up to $4000\Delta$.



\begin{figure*}[t]
\centering
\includegraphics[width=1.0\textwidth]{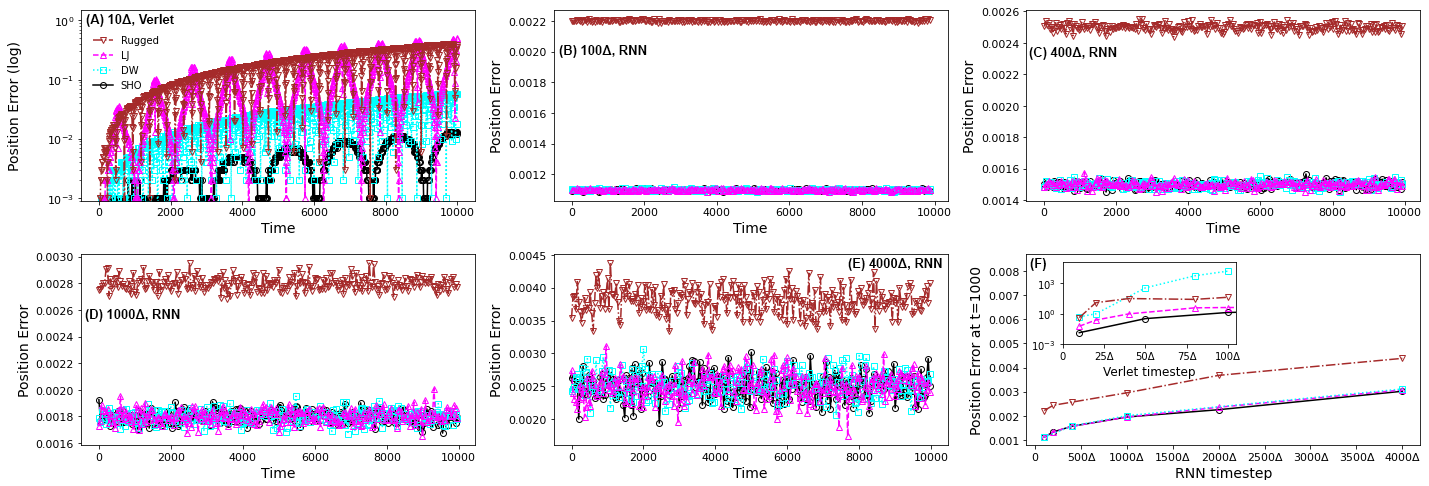}
\caption{
Errors incurred in the predictions of particle positions made by Verlet integrator and four RNN operators for four 1D systems. 
Black circles, blue squares, magenta up triangles, and brown down triangles represent errors in position predictions for a particle in a simple harmonic potential, double well potential, Lennard-Jones potential, and rugged potential respectively (see the main text for a detailed description of these 1D systems).
(A) shows errors (log scale) as a function of time for dynamics extracted sing the Verlet integrator with timestep $dt = 10\Delta$.
(B), (C), (D) and (E) show 
errors as a function of time when the time evolution is performed by the RNN operators using timestep $\Delta_{\mathrm{R}} = 100\Delta$, $400\Delta$, $1000\Delta$ and $4000\Delta$ respectively.
(F) shows errors in position predictions by the RNN operators at time $t=1000$ as a function of timestep for the same four 1D systems (inset shows results on a log scale obtained using the Verlet integrator). 
}
\label{fig:manypotentials}
\end{figure*}

\subsection{One particle systems in 1D}
\label{sec:1Dresults}

Our first set of experiments focus on training and testing the RNN-based $\mathscr{R}$ operators to predict the dynamics of one particle systems in 1D. Results are shown for dynamics predicted by four $\mathscr{R}$ operators for four representative one particle systems, each characterized with a different 1D potential. 
For each of these one particle systems in 1D, the training and validation datasets are generated by recording the dynamics associated with a few discrete initial configurations, particle masses, and, in some cases, parameters characterizing the potential energy. We describe the process in detail for the case of a particle in a simple harmonic potential $U=1/2kx^2$. A similar process is followed for all other 1D systems (see Appendix for details).

For the 1D system of a particle in a simple harmonic potential, the dataset consists of ground-truth trajectories associated with input systems generated by sweeping over 
5 discrete values of initial position of the particle ($x_0 = -10, -8, -6, -4, -2$), 10 discrete values of particle mass ($m = 1, 2, \ldots, 9, 10$), and 10 discrete values of spring constant ($k = 1, 2, \ldots, 9, 10$).
The trajectory data for each of these input systems is obtained using simulations performed with the Verlet integrator with $\Delta=0.001$ up to time $t=200$. This process generates a dataset of 500 simulations, each having 400,000 position and velocity values. 
The entire dataset is randomly shuffled and separated into training and validation sets using a ratio of 0.8:0.2. In other words, $80\%$ of the simulations (400 systems) are selected randomly as part of the training dataset, and the remaining $20\%$ (100 systems) are separated into the validation dataset. 
The testing dataset to evaluate the predictions of the RNN-based operator comprises 100 input systems characterized with $m, x_0, k$ values distinct from those used in the input systems in the training and validation datasets, including many combinations of these 3 parameters that lie outside the boundary of the input domains described above.
For experimental evaluation of the operator $\mathscr{R}$, systems characterized with input parameters outside the boundary of the ranges associated with the training and validation datasets are randomly selected from the testing dataset. These systems provide a more challenging task for the operator compared to the systems within the training ranges. The same process is followed for the other 3 one particle systems.

\begin{figure*}[t]
\centering
\includegraphics[width=1.0\textwidth]{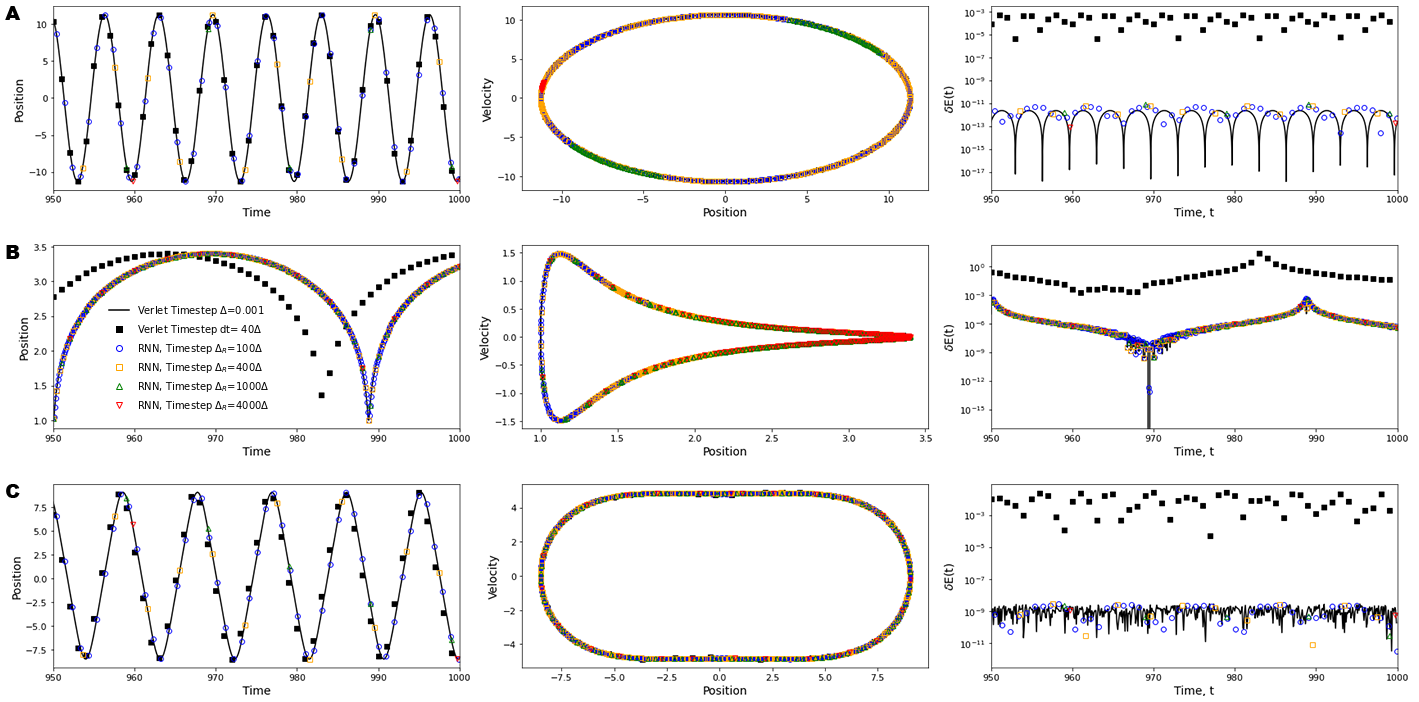}
\caption{
Time evolution performed by the RNN operators for a single particle in three 1D potentials from time $t=950$ to $1000$. Results are presented as position vs time (left column), velocity vs position (middle column), and energy deviation vs time (right column).
Open symbols are dynamics predicted by the RNN operators with timestep $\Delta_{\mathrm{R}} = 100\Delta$ (blue circles), $400\Delta$ (orange squares), $1000\Delta$ (green up triangles) and $4000\Delta$ (red down triangles). 
Solid black squares are dynamics produced by the Verlet integrator with timestep $50\Delta$. 
(A) Dynamics of a particle of mass $m=14.4$ and initial position $x_0=-11.3$ in a simple harmonic potential with spring constant $k=12.8$. (B) Dynamics of a particle of mass $m=0.9$ and initial position $x_0=3.4$ in a Lennard-Jones potential. (C) Dynamics of a particle of mass $m=8.0$ and initial position $x_0=-8.5$ in a rugged potential. 
RNN predictions are in excellent agreement with the ground truth results (black lines) obtained using the Verlet integrator with timestep $\Delta=0.001$. 
}
\label{fig:3potentials-energy}
\end{figure*}

Figure \ref{fig:manypotentials} shows the errors incurred in the predictions of particle positions made by 4 $\mathscr{R}$ operators for 4 one particle 1D systems: a particle of mass $m=14.4$ and initial position $x_0=-11.3$ in a simple harmonic potential $U(x) = 1/2 k x^2$ with $k=12.8$, a particle of mass $m=13$ and initial position $x_0=-12.5$ in a double well potential $U(x) = x^4 / 4 - x^2 / 2$, a particle of mass $m=13.2$ and initial position $x_0=3.4$ in a Lennard-Jones potential $U(x) = 4 \left(1/x^{12}  - 1/x^6\right)$, and a particle of mass $m=11.1$ and initial position $x_0=-8.5$ in a rugged potential $U(x) = 1/50\left(x^4 - x^3 - 16x^2 + 4x + 48 + 10\sin{\left(30x+150\right)}\right)$.
Figures \ref{fig:manypotentials} B, C, D and E show the errors incurred in the predictions of particle positions as a function of time $t$ for timestep $\Delta_{\mathrm{R}} = 100\Delta$, $400\Delta$, $1000\Delta$ and $4000\Delta$ respectively. These trajectory errors are computed as $\delta r (t) = \left| \vec{r}(t; \Delta_{\mathrm{R}}) - \vec{r}_{V}(t, \Delta) \right|$, where $\vec{r}(t; \Delta_{\mathrm{R}})$ is the position vector of the particle at time $t$ predicted by the RNN operator $\mathscr{R}$ with timestep $\Delta_{\mathrm{R}}$ and $\vec{r}_{V}(t; \Delta)$ is the corresponding ground truth result produced by the Verlet integrator with timestep $\Delta = 0.001$.
For all 4 one particle 1D systems, the errors $\delta r (t)$ are $O( 10^{-3})$ for all $\Delta_{\mathrm{R}}$ values and do not increase with time $t$ up to 10000, even for $\Delta_{\mathrm{R}}$ as large as $4000\Delta$. 
In stark contrast, for the same systems, Figure \ref{fig:manypotentials}A shows that the trajectory errors associated with time evolution performed using the Verlet integrator with a timestep of $10\Delta$, $\delta r (t) = \left| \vec{r}_V(t; 10\Delta) - \vec{r}_{V}(t; \Delta) \right|$, increase exponentially with time, reaching values as high as $O( 10^{-1})$.

Errors in position predictions made by the $\mathscr{R}$ operators rise with increasing the complexity of the 1D potential. For example, for all $\Delta_{\mathrm{R}}$, trajectory errors are higher for the 1D system of a particle in a rugged potential compared to the 1D system of a particle in a simple harmonic potential. 
Figure \ref{fig:manypotentials}F shows the errors in position predictions at time $t=1000$ as a function of timestep for the same four 1D systems. In each case, the errors increase as the timestep $\Delta_{\mathrm{R}}$ is increased from $100\Delta$ to $4000\Delta$, but remain within an order of magnitude $O(10^{-3})$. 
On the other hand, errors incurred in positions evolved using the Verlet integrator (Figure \ref{fig:manypotentials}F inset) rise steeply with increasing timestep and are three orders of magnitude larger for a timestep of $100\Delta$ compared to the errors associated with predictions made by the $\mathscr{R}$ operators. 

Figure \ref{fig:3potentials-energy} shows the predictions made by the RNN operators for positions, velocities, and energy deviations associated with the dynamics of a particle in three 1D potentials: particle of mass $m=14.4$ and initial position $x_0=-11.3$ in a simple harmonic potential with $k=12.8$ (A), particle of mass $m=0.9$ and initial position $x_0=3.4$ in an LJ potential (B), and particle of mass $m=8.0$ and initial position $x_0=-8.5$ in a rugged potential (C). 
Results for each system are presented in three graphs: position vs time, velocity vs position, and energy deviation vs time. Energy deviation $\delta E_t$ is defined as $\delta E_t = |E_t-E_0|/|E_0|$, where $E_t$ and $E_0$ are the total energy of the system at time $t$ and the initial time $t=0$, respectively.

For each 1D system, the positions and velocities predicted by the associated $\mathscr{R}$ operator using timestep $\Delta_{\mathrm{R}} = 100\Delta$, $400\Delta$, $1000\Delta$ and $4000\Delta$ are in excellent agreement with the ground truth results. The associated energy deviation $\delta E_t$ tracks the ground truth energy deviation for all values of  $\Delta_{\mathrm{R}}$.
On the other hand, positions and velocities produced by the Verlet integrator with a timestep of $40\Delta$ ($<\Delta_{\mathrm{R}}$) show large deviations from the ground truth for all three 1D systems; the corresponding energy deviations are also orders of magnitude larger compared to the results obtained with RNN operators and the ground truth results.



\begin{figure*}[t]
\centering
\includegraphics[width=\textwidth]{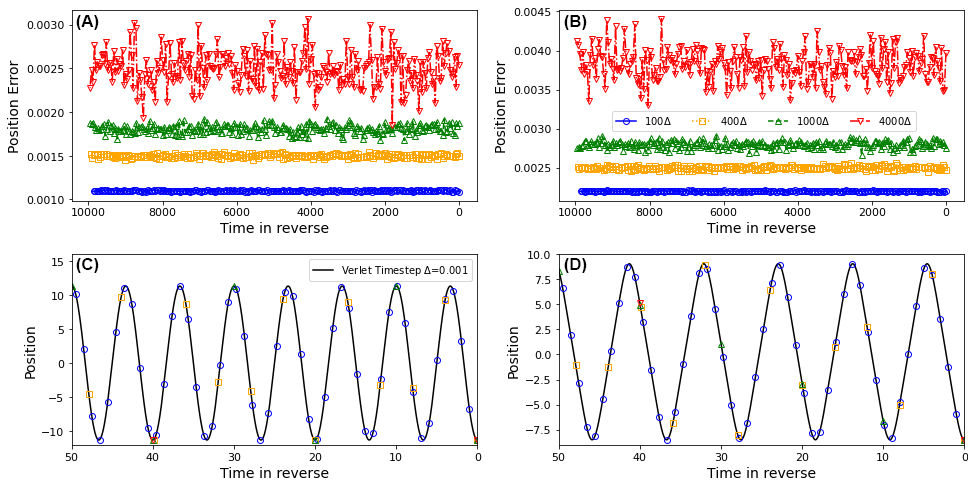}
\caption{
Trajectory errors and particle positions predicted during the backward time evolution performed by the RNN operators for two 1D systems: a particle of mass $m=14.4$ and initial position $x_0=-11.3$ in a simple harmonic potential with $k=12.8$ (A and C), and a particle of mass $m=11.1$ and initial position $x_0=-8.5$ in a rugged potential (B and D). 
(A) and (B) show the errors incurred in positions predicted vs time $t$ from $t=10000$ to $t=0$ for backward time evolution by the RNN operators with timestep 
$\Delta_{\mathrm{R}} = 100\Delta$ (circles), $400\Delta$ (squares), $1000\Delta$ (up triangles), and $4000\Delta$ (down triangles).
(C) and (D) show the corresponding positions as a function of $t$ for the two 1D systems from $t=50$ to $t=0$. RNN predictions are in excellent agreement with the ground truth results (black lines) obtained using the Verlet integrator with timestep $\Delta$. }
\label{fig:time_reversal}
\end{figure*}

In addition to the forward time evolution, we find that the RNN operators can accurately perform backward time evolution of 1D systems for an arbitrary length of time by utilizing the trajectory data in reverse order without undergoing any re-training using the time-reversed trajectories. 
Analogous to the process followed for the forward time evolution, we feed a sequence of length $S_R=5$ of the future states of the trajectory starting at an arbitrary time $t + S_R\Delta_{\mathrm{R}}$, and predict the state at time $t -  \Delta_{\mathrm{R}}$. The backward evolution terminates with the prediction at $t = 0$. 
Representative results for the backward time evolution are shown in Figure \ref{fig:time_reversal} for a particle of mass $m=14.4$ and initial position $x_0=-11.3$ in a simple harmonic potential with $k=12.8$ (A and C) and a particle of mass $m=11.1$ and initial position $x_0=-8.5$ in a rugged potential (B and D). 
Figures \ref{fig:time_reversal}A and \ref{fig:time_reversal}B show that the $\mathscr{R}$ operators generate accurate backward time evolution of these two systems starting from $t=10000$ to $t=0$ for $\Delta_{\mathrm{R}} = 100\Delta,400\Delta,1000\Delta,4000\Delta$. Errors in position predictions are $O( 10^{-3})$ for all $\Delta_{\mathrm{R}}$, similar to the errors in the forward trajectory evolution predicted by the same operators (Figure \ref{fig:manypotentials}), and do not increase as time evolves backward.
Figure \ref{fig:time_reversal}C and \ref{fig:time_reversal}D show the predicted positions vs time in reverse for the two systems from $t=80$ to $t=0$ for $\Delta_{\mathrm{R}} = 100\Delta,400\Delta,1000\Delta,4000\Delta$. RNN predictions are in excellent agreement with the ground truth results obtained using Verlet integrator with timestep $\Delta$. 
In addition to exhibiting the time-reversal symmetry, we find that the RNN operators, with no explicit training to satisfy the symplectic condition, approximately preserve the symplectic property for timesteps up to $1000\Delta$ (see Appendix for details).


\begin{figure*}[t]
\centering
\includegraphics[width=1.0\textwidth]{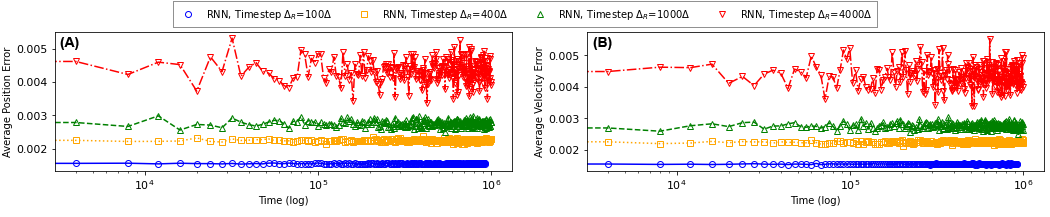}
\caption{
Average errors as a function of time $t$ (log scale) associated with the predictions of the RNN operator for the positions (A) and velocities (B) of a 3D system of $16$ particles interacting via Lennard-Jones forces under periodic boundary conditions.
Results are shown for the time evolution performed using timestep
$\Delta_{\mathrm{R}} = 100\Delta$ (circles), $400\Delta$ (squares), $1000\Delta$ (up triangles) and $4000\Delta$ (down triangles).
For all $\Delta_{\mathrm{R}}$, the errors are $O(10^{-3})$ during the entire evolution up to $t=10^6$. 
}
\label{fig:manyparticle}
\end{figure*}

\begin{figure*}[t]
\centering
\includegraphics[scale=0.45]{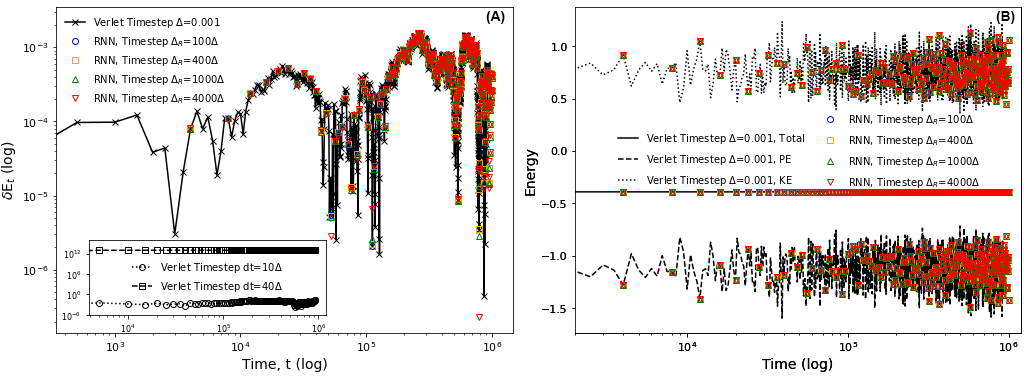}
\caption{
(A) Energy deviation $\delta E_t$ (defined in the main text) as a function of time $t$ (log scale) associated with the dynamics predicted by the RNN operator for the same 3D system of 16 particles as described in Figure \ref{fig:manyparticle}. 
Results are shown for the time evolution performed using timestep
$\Delta_{\mathrm{R}} = 100\Delta$ (circles), $400\Delta$ (squares), $1000\Delta$ (up triangles) and $4000\Delta$ (down triangles).
Inset shows the corresponding results using the Verlet integrator with timesteps $10\Delta$ and 40$\Delta$.
For all values of $\Delta_{\mathrm{R}}$, the dynamics generated by the RNN operator track the ground truth result (black crosses) produced using the Verlet integrator with timestep $\Delta=0.001$. 
(B) The total, potential, and kinetic energies associated with the dynamics predicted by the RNN operator using the same four $\Delta_{\mathrm{R}}$ values; symbols have the same meaning as in (A). Lines represent the ground truth results obtained using the Verlet integrator for total (solid), potential (dashed), and kinetic (dotted) energies. For all $\Delta_{\mathrm{R}}$, the predicted energy profiles track the ground truth results.
}
\label{fig:manyparticle-energy-profile}
\end{figure*}

\subsection{Few particle systems in 3D}
\label{sec:3Dresults}

Our next set of experiments focus on training and testing the RNN-based $\mathscr{R}$ operators to predict the dynamics of few particles in 3D. Three separate $\mathscr{R}$ operators are designed to predict the dynamics of three few-particle systems with $N=3$, $N=8$, and $N=16$ particles interacting via shifted and truncated Lennard-Jones (LJ) potentials in a cubic box with periodic boundary conditions. All particles have the same mass $m=1$ and interact with the following LJ pair interaction potential:
\begin{eqnarray*} \label{eq:data11}
U(r) &= 4\epsilon \left(\left(\frac{1}{r}\right)^{12} - \left(\frac{1}{r}\right)^{6}\right) + 0.0163 \epsilon \quad \mathrm{for} \quad r \le 2.5, \\
&= 0 \quad \mathrm{for} \quad r > 2.5. 
\end{eqnarray*}

For each of the three $N-$particle cases, the training and validation datasets consist of ground-truth trajectories produced by simulations of 5000 systems. These systems are generated by selecting different initial positions $\vec{r}_0$ for the $N$ particles.
The process begins by randomly selecting each of the three Cartesian coordinates $x_0, y_0, z_0$ of one particle between $-3.0$ and $3.0$, and then placing all other particles next to the initial seed particle with a step size of $0.3508$, ensuring that there are no particle overlaps and all particles have Cartesian coordinates between $-3.0$ and $3.0$. 
In all simulations used to create the training and validation datasets, the initial velocities for all particles are set to $0$, and the characteristic LJ energy $\epsilon$ is set to $1$. Simulations are performed using the Verlet integrator with $\Delta=0.001$ up to time $t=2000$. 

In each case, the entire dataset is randomly shuffled and separated into training and validation sets using a ratio of 0.8:0.2. For example, in the case of $N=16$ particle system, $80\%$ of the simulations (4000 systems) are randomly selected to form the training dataset, and the remaining $20\%$ (1000 systems) are separated into the validation dataset. 
For experimental evaluation of the RNN operators, separate testing datasets for $N=3$, $N=8$, and $N=16$ particle systems are generated using simulations of particles of mass $m = 1$ performed up to time $t=1,000,000$. 
In these simulations, the three Cartesian coordinates associated with the initial positions of all the particles are randomly selected between $-3.0$ and $3.0$ ensuring no overlapping particles.
The initial particle velocities are sampled from a Boltzmann distribution with a reduced temperature of 1, and the characteristic LJ energy $\epsilon$ is set to $2$. Thus, the RNN operators are tasked to make predictions for test samples that are very different from the typical sample in the training and validation datasets (representative energy profiles associated with typical samples in test and training datasets are shown in Figure \ref{fig:manyparticle-Train-energy-profile} in Appendix).

\begin{table*}[t]
\centering
\caption{Speedup $S$ for time evolution by the RNN operators using timestep $\Delta_{\mathrm{R}}$ as shown in the column heading.}
\label{tab.ml.speedups}
\begin{tabular}{lcrrrrr}
\textbf{Experiment} & $100\Delta$ & $200\Delta$ & $400\Delta$ & $1000\Delta$ & $2000\Delta$ & $4000\Delta$ \\
\hline
\textbf{1D, Simple Harmonic} & 0.5 & 1.3 & 3.2 & 8.6 & 20.0 & 45.0 \\
\textbf{1D, Double Well} & 0.6 & 1.2 & 2.8 & 8.7 & 17.3 & 38.0 \\
\textbf{1D, Lennard-Jones} & 0.9 & 1.5 & 3.9 & 12.8 & 22.5 & 42.3 \\
\textbf{1D, Rugged} & 0.4 & 0.8 & 2.1 & 4.7 & 9.7 & 20.6 \\
\textbf{3D, 8 particles} & 600 & 1000 & 1500 & 5500 & 8300 & 12000 \\
\textbf{3D, 16 particles} & 3000 & 4900 & 7100 & 20000 & 28000 & 32000\\
\hline
\end{tabular}

\end{table*}

For all the three $N$-particle systems, we find that the associated $\mathscr{R}$ operators accurately evolve the positions and velocities of the particles with timestep $\Delta_{\mathrm{R}}$ as large as $4000\Delta$, yielding energy-conserving dynamics up to time $t = 10^6$.
In the interest of brevity, we discuss the results for the 3D system with $N=16$ particles.
Figures \ref{fig:manyparticle}A and \ref{fig:manyparticle}B show the average error associated with the RNN predictions for the positions and velocities of $N=16$ particles as a function of time respectively. 
The average error in the prediction of positions is computed as $\delta r (t) = \sum_{i=1}^{N} \left| \vec{r}_{i}(t) - \vec{r}_{i, V}(t) \right| / N$, where $\vec{r}_i(t)$ is the 3D position vector of the $i^{\mathrm{th}}$ particle at time $t$ predicted by the RNN operator $\mathscr{R}$ with timestep $\Delta_{\mathrm{R}}$ and $\vec{r}_{i, V}(t)$ is the corresponding ground truth result at the same time $t$ produced by the Verlet integrator with timestep $\Delta = 0.001$. 
The average error in the prediction of velocities is computed as $\delta v (t) = \sum_{i=1}^{N} \left| \vec{v}_{i}(t) - \vec{v}_{i, V}(t) \right| / N$, where $\vec{v}_i(t)$ is the 3D velocity vector of the $i^{\mathrm{th}}$ particle at time $t$ predicted by the RNN operator $\mathscr{R}$ with timestep $\Delta_{\mathrm{R}}$ and $\vec{v}_{i, V}(t)$ is the corresponding ground truth result at the same time $t$ produced by the Verlet integrator with timestep $\Delta = 0.001$. 
Results are shown for time evolution performed by $\mathscr{R}$ for $\Delta_{\mathrm{R}}=100\Delta, 400\Delta, 1000\Delta$, and $4000\Delta$.
For all values of $\Delta_{\mathrm{R}}$, we find that the errors $\delta r(t)$ and $\delta v(t)$ are small up to $t=10^6$. These errors rise as $\Delta_{\mathrm{R}}$ increases but remain $O(10^{-3})$ for the entire duration of the time evolution.

Figure \ref{fig:manyparticle-energy-profile}A shows the energy deviation $\delta E_t = |E_t-E_0|/|E_0|$ associated with the time evolution of the 3D system of 16 particles predicted by the RNN operator $\mathscr{R}$ using $\Delta_{\mathrm{R}} = 100\Delta, 400\Delta, 1000\Delta$, and $4000\Delta$. $E_t$ and $E_0$ are the total energy of the system at time $t$ and the initial time $t=0$, respectively. 
For all values of $\Delta_{\mathrm{R}}$, the dynamics generated by the RNN operator exhibits excellent energy conservation: $\delta E_t$ $\lesssim 10^{-3}$ for up to $t=10^6$ and tracks the ground truth result produced using the Verlet integrator with timestep $\Delta$.  
In stark contrast, the dynamics produced using Verlet integrator with a timestep of $40\Delta$ (inset in Figure \ref{fig:manyparticle-energy-profile}A) exhibits a rapid energy divergence with $\delta E_t \sim 10^{12}$ for $t>10^2$.
Figure \ref{fig:manyparticle-energy-profile}B shows the kinetic, potential, and total energies associated with the dynamics predicted by the RNN operator using timestep $\Delta_{\mathrm{R}} = 100\Delta, 400\Delta, 1000\Delta$, and $4000\Delta$. The corresponding ground truth results obtained with the Verlet integrator using timestep $\Delta=0.001$ are also shown. For all $\Delta_{\mathrm{R}}$, the total energy predicted by $\mathscr{R}$ as a function of time is conserved. All the predicted energy profiles track the ground truth results up to $t=10^6$. 

\subsection{Performance Enhancement}


We now discuss the performance enhancement resulting from using the deep learning approach presented here to perform simulations of one-particle and few-particle systems. 
For a given system, our approach uses the Verlet integrator to kickstart the simulation and the RNN operator $\mathscr{R}$ designed for that system to evolve the dynamics forward in time. Incorporating this detail, we propose the following speedup metric $S$ to quantify the performance enhancement:
\begin{equation}
S = \frac{S_{\mathrm{T}}t_{\mathrm{V}}}{S_{\mathrm{V}} t_{\mathrm{V}} + \left(S_{\mathrm{T}} - S_{\mathrm{V}}\right) t_{\mathrm{R}} \Delta / \Delta_{\mathrm{R}}}
\end{equation}
where $S_{\mathrm{T}}$ is the total number of steps needed if the time evolution is performed using only the Verlet integrator and $S_{\mathrm{V}} = \Delta_{\mathrm{R}}(S_{\mathrm{R}} - 1)/\Delta$ is the total number of steps that generate the initial trajectories using Verlet to kickstart the simulation. $t_{\mathrm{V}}$ and $t_{\mathrm{R}}$ are the times for one forward step propagation using Verlet and $\mathscr{R}$ respectively. 
In the speedup $S$, we have not accounted for the time spent on creating training and validation datasets, which is a one-time investment of $<24$ hours for the experiments shown.
$S$ is 1 if $S_{\mathrm{T}}=S_{\mathrm{V}}$, that is, when no time evolution is performed using the RNN operator $\mathscr{R}$. 
In the limit $S_{\mathrm{T}} \gg S_{\mathrm{V}}$, we obtain $S \approx t_{\mathrm{V}}\Delta_{\mathrm{R}} / (t_{\mathrm{R}}\Delta)$. Clearly, the greater the ratio $\Delta_{\mathrm{R}}/\Delta$, the higher the speedup.

Table \ref{tab.ml.speedups} shows the speedup $S$ for the time evolution by the RNN operators using different timestep $\Delta_{\mathrm{R}}$. Results are shown for 1D experiments (first 4 rows) and 3D experiments (last 2 rows), and for $\Delta_{\mathrm{R}} = 100, 200, 400, 1000, 2000,$ and $4000$. 
In all cases, $S$ is  computed for time evolution up to $t=10^6$ with $S_T=10^9$ steps.
We find that the time $t_{\mathrm{V}}$ for one forward step propagation using Verlet varies by four orders of magnitude across the different experiments, ranging from $\approx 9\times 10^{-6}$ seconds (for the 1D system with simple harmonic potential) to $4\times 10^{-2}$ seconds (for the 3D system with 16 particles).
In contrast, the time $t_{\mathrm{R}}$ for one forward step propagation using the different RNN operators varies by only one order of magnitude across experiments, ranging from $\approx 3\times 10^{-4}$ seconds (for the 1D system with simple harmonic potential) to $\approx 2 \times 10^{-3}$ seconds (for the 3D system with 16 particles).

We find that the speedup $S>1$ for most experiments, signaling an enhancement in performance when the time evolution is performed using our deep learning approach. Low $S < 1$ values, observed mostly for the time evolution of the 1D systems with $\Delta_{\mathrm{R}} = 100\Delta$, can be attributed to the relatively large time for one forward step propagation using RNN ($t_\mathrm{R} \gg t_{\mathrm{V}}$). 
As expected, $S$ rises with increasing $\Delta_{\mathrm{R}}$. 
The largest values of $S$ are recorded for 3D systems with more number of particles. 
In these cases, large increases in $S$ result from both the use of large timestep $\Delta_{\mathrm{R}}$ and the small time associated with the forward step propagation using RNN operators ($t_{\mathrm{R}} < t_{\mathrm{V}}$) . 
For example, in the case of the time evolution of the 3D system of 16 particles with $\Delta_{\mathrm{R}}=4000\Delta$, we find that $t_{\mathrm{R}} \approx 0.0026$ seconds is an order of magnitude smaller than $t_{\mathrm{V}}\approx 0.04392$ seconds, resulting in the speedup $S \approx 32000$.

\subsection{Limitations and Outlook}

%
We now explore the limits of the applicability of our deep learning approach.
All RNN operators are trained using ground-truth trajectories associated with systems generated by sweeping input parameters over a finite range of values. Our results demonstrate that these operators can successfully perform time evolution of unseen input systems characterized with parameters that lie within and outside the ranges associated with the training datasets. 
However, as the input systems become progressively different from the systems in the training datasets, e.g., by selecting parameters that are well beyond the parameter ranges associated with the training datasets, we expect the RNN operators to produce inaccurate time evolution, generating trajectories that deviate from the ground truth results.
    
Consider the 1D case of one particle in a simple harmonic potential for which we trained the RNN operator with ground-truth trajectories associated with input systems generated by sweeping over 5 discrete values of the initial position $x_0$ of the particle in the range $x_0 \in [-10,-2]$, 10 discrete values of particle mass $m$ in the range $m \in [1,10]$, and 10 discrete values of spring constant $k$ in the range $k \in [1,10]$. We extrapolated to an input system characterized with parameters $x_0 = -11.3, m = 14.4, k = 12.8$ in order to test the predictions of the RNN operator. For this test system, the operator produced accurate energy-conserving time evolution (Section \ref{sec:1Dresults}). However, for a particle with the same initial position $x_0=-11.3$ and mass $m=14.4$, the trajectories predicted by the RNN operator become progressively inaccurate as the spring constant $k$ is increased to values greater than $2\times$ the maximum $k$ value used in the training process (i.e., for $k\gtrsim20$).

\begin{figure*}[th]
\centering
\includegraphics[scale=0.40]{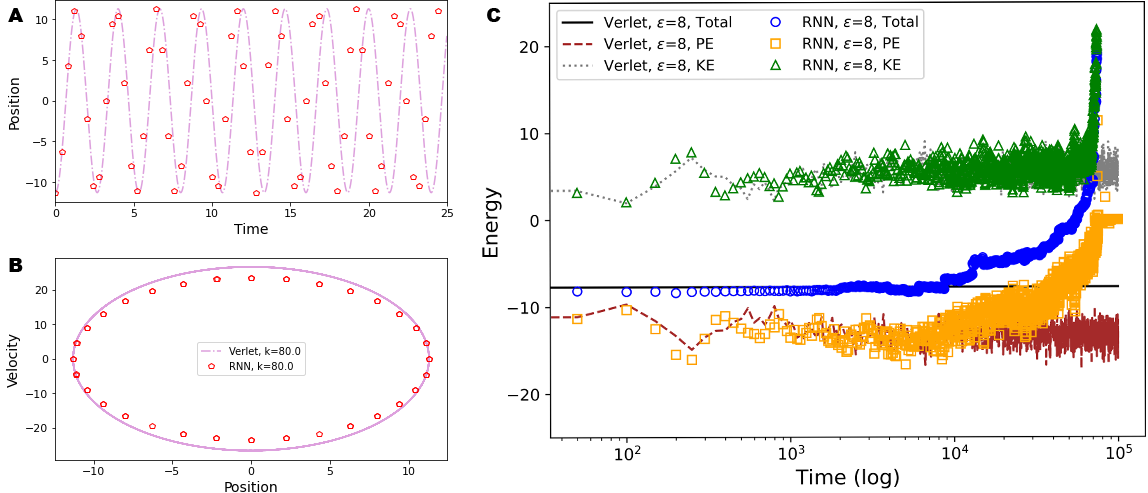}
\caption{
Failure cases illustrating the limits of the applicability of the deep learning approach. 
(A) and (B) show the time evolution predicted by the RNN operator with a timestep $\Delta_{\mathrm{R}} = 100\Delta$ (open circles) for a particle of mass $m=14.4$ and initial position $x_0=-11.3$ in a 1D simple harmonic potential characterized with spring constant $k=80.0$. Dynamics are shown as position vs time (A) and velocity vs position (B) plots from time $t=0$ to $t=25$. 
The RNN predictions quickly deviate from the ground truth results (lines) obtained using the Verlet integrator with timestep $\Delta = 0.001$. 
(C) shows the energies associated with the time evolution predicted by the RNN operator with a timestep $\Delta_{\mathrm{R}} = 100\Delta$ (symbols) for the 3D system of the 16 particles interacting via Lennard-Jones potentials characterized with energy parameter $\epsilon=8$. The total, potential, and kinetic energies are shown from time $t=0$ to $t=10^6$. All predicted energies start deviating quickly from the ground truth results (lines) obtained using the Verlet integrator with timestep $\Delta = 0.001$. The deviations get progressively larger with increasing time $t$.
}
\label{fig:limitations}
\end{figure*}

Figures \ref{fig:limitations}A and \ref{fig:limitations}B illustrate a failure case by showing the time evolution for a particle of mass $m=14.4$ and initial position $x_0=-11.3$ in a simple harmonic potential characterized with spring constant $k=80.0$ (which is $8\times$ the maximum $k$ value used in the training dataset). The time evolution by the RNN operator uses a timestep $\Delta_{\mathrm{R}} = 100\Delta$, and results are shown from $t=0$ to $t=25$. 
After a small duration of time $t > 1$, the RNN predictions for positions and velocities deviate from the ground truth results obtained using the Verlet integrator. Recall that the same RNN operator predicted accurate time evolution up to $t=1000$ for this 1D system with $k=12.8$ (Section \ref{sec:1Dresults}). Similar limits in extrapolation and generalizability are observed for other 1D systems.

We next consider the 3D case of 16 particles interacting with Lennard-Jones (LJ) potentials in periodic boundary conditions. For this case, the RNN operator was trained with ground-truth trajectories associated with input systems generated by sweeping over many discrete values of the initial positions of the particles. Simulations of all systems in the training dataset were initialized with zero particle velocities and the LJ interactions between particles were characterized with energy parameter $\epsilon = 1$. To test the predictions of the RNN operator, we extrapolated to an input system for which the velocities of the 16 particles were sampled from a Boltzmann distribution with a reduced temperature of 1, and the LJ interactions were characterized with the energy parameter $\epsilon = 2$. For this test system, the operator produced accurate energy-conserving time evolution (Section \ref{sec:3Dresults}). However, we find that the time evolution predicted by the RNN operator becomes progressively inaccurate as $\epsilon$ is increased to values over $5$.
    
Figure \ref{fig:limitations}C illustrates a failure case by showing the energy profiles associated with the time evolution performed by the RNN operator for 16 particles interacting via LJ potentials characterized with $\epsilon=8$. Simulation is initialized with randomly selected positions and with velocities sampled from the Boltzmann distribution at a reduced temperature of 1. The time evolution by the RNN operator uses a timestep $\Delta_{\mathrm{R}} = 100\Delta$, and results are shown from $t=0$ to $t=10^6$. All energies (kinetic, potential, and total) start deviating from the ground truth results right from the beginning, the deviations getting progressively stronger with increasing time $t$. As $t$ increases to values beyond $10^4$, the total energy starts to diverge. Recall that the same RNN operator predicted accurate time evolution from $t=0$ to $t=10^6$ for this 3D system with $\epsilon=2$ (Section \ref{sec:3Dresults}). 
For both one particle systems in 1D and few particle systems in 3D, addressing the failure cases will involve expanding the range of input parameters characterizing the systems and including the associated trajectory data in the training of the RNN operators.  
    
In our current formulation, the RNN operators are designed and trained to furnish the time evolution of systems described by the selected potential energy describing the particles. Thus, the RNN operator trained to learn the dynamics of one potential energy landscape (e.g., one particle in a simple harmonic potential) is, by design, not capable to predict the dynamics of another closely related but qualitatively different potential energy landscape (e.g., one particle in a double well potential). 
For a complex potential energy landscape with multiple basins and barriers, the associated RNN operator will need to ``see'' a diverse group of trajectories corresponding to different regions of the energy landscape in order to accurately furnish the time evolution. The complexity of the energy landscape may require changes in the architectural configuration of the RNN operators and may also lead to longer training times. Similarly, the RNN operators will need to be trained with ground truth trajectories associated with different representative assembly behaviors (e.g., phase changes in 3D systems of many particles interacting with LJ potentials) in order for them to successfully evolve the dynamics for corresponding thermodynamic statepoints.
    
Our future work will explore ways to enhance the generalizability of the RNN operators and extend the applicability of our deep learning approach to systems described with complex energy landscapes. 
In this initial study, we have limited our focus on few-particle systems and one type of thermodynamic ensemble. Future work will also focus on scaling the approach to a larger number of particles and testing the accuracy of the RNN operators in different thermodynamic ensembles.

\section{Conclusion}

We have introduced a deep learning approach that utilizes recurrent neural networks (RNNs) to design operators that solve Newton's equations of motion and evolve the dynamics of particles by utilizing timesteps orders of magnitude larger than the typical timestep used in numerical integrators such as Verlet. We have obtained state-of-the-art results in terms of the timesteps, the number of particles, and the complexity of the potential characterizing the interactions between particles.
The RNN operators learn both the interaction potentials and the dynamics of the particles based on their experience with the ground-truth solutions of Newton's equations of motion.
These operators produce accurate predictions for the time evolution of particles accompanied with excellent energy conservation over a variety of force fields using up to $4000 \times$ larger timestep than the Verlet integrator.
The use of deep learning methods in tasks central to molecular dynamics simulations is a critical first step towards the long-term goal of machine-learning-assisted molecular dynamics simulations of many particle systems. 
Further, the idea of formulating the dynamics of particles into a sequence processing problem solved via the use of recurrent neural networks illustrates an important approach to learn the time evolution operators, which is applicable across different fields including fluid dynamics and robotics \cite{bar2019learning,shen2017essential,kates2019predicting}.

 \section*{Acknowledgments}
 This work is partially supported by the NSF through awards 1720625 and DMR-1753182, and by the DOE through award DE-SC0021418. G.C.F was partially supported by NSF CIF21 DIBBS 1443054 and CINES 1835598 awards. V.J. thanks M. O. Robbins for useful discussions.


\section*{APPENDIX}

\subsection*{Training and Validation Dataset Preparation for 1D Systems}

\begin{figure}[t]
\centering
\includegraphics[width=0.48\textwidth]{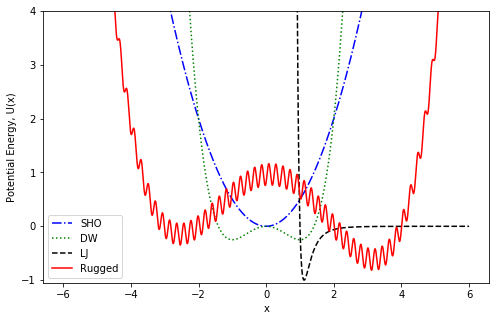}
\caption{Potential energies associated with the 1D experiments. Dash-dotted, dotted, dashed, and solid lines represent simple harmonic (SHO), double well (DW), Lennard-Jones (LJ), and rugged potentials respectively.}
\label{fig:potential_plots}
\end{figure}

Here, we describe the datasets used for training the RNN operators to predict the dynamics of 1D one-particle systems. The potential energy functions characterizing the four 1D systems are shown in Figure \ref{fig:potential_plots}.
In each case, the Verlet integrator with a timestep $\Delta = 0.001$ is used to generate the ground truth trajectories for up to time $t = 200$.
For all four 1D systems described below, the testing dataset to evaluate the predictions of the associated RNN-based operator comprises 100 input systems characterized with parameters distinct from those used in the input systems in the training and validation datasets. These 100 test systems also include systems characterized with all parameters lying outside the boundary of the parameter ranges associated with systems in the training and validation datasets.

\paragraph{Particle in a simple harmonic potential}
For this system, the potential energy is given by
\begin{equation} \label{eq:data1}
U = \frac{1}{2} k x^2,
\end{equation}
where $k$ is the spring constant. 
The training and validation datasets consist of ground-truth trajectories associated with simulations of 500 input systems generated by sweeping over 5 discrete values of initial position of the particle $x_0 = -10, -8, -6, -4, -2$; 10 discrete values of particle mass $m = 1, 2, \ldots, 9, 10$; and 10 discrete values of spring constant $k = 1, 2, \ldots, 9, 10$. Each simulation produces a trajectory with 400,000 position and velocity values. 

\paragraph{Particle in a double well potential}
For this system, the potential energy is given by
\begin{equation} \label{eq:data7}
U = \frac{1}{4}x^4 - \frac{1}{2}x^2.
\end{equation}
The training and validation datasets consist of ground-truth trajectories associated with simulations of 500 input systems generated by sweeping over 10 discrete values of particle mass $m = 1, 2, \ldots, 9, 10$; and 50 uniformly-distributed discrete values of initial position of the particle $x_0 \in [-3.1, 3.1]$. Each simulation produces a trajectory with 400,000 position and velocity values. 

\paragraph{Particle in a Lennard-Jones potential}
For this system, the potential energy is given by
\begin{equation} \label{eq:data5}
U(x) = 4\left(\left(\frac{1}{x}\right)^{12} - \left(\frac{1}{x}\right)^{6}\right).
\end{equation}
The training and validation datasets consist of ground-truth trajectories associated with simulations of 500 input systems generated by sweeping over 10 discrete values of particle mass $m = 1, 2, \ldots, 9, 10$; and 50 uniformly-distributed discrete values of initial position of the particle $x_0 \in [1.0, 3.0]$. Each simulation produces a trajectory with 400,000 position and velocity values. 

\paragraph{Particle in a rugged potential}
For this system, the potential energy \cite{wang2019machine} is given by 
\begin{equation} \label{eq:data9}
U(x) = \frac{x^4 - x^3 - 16x^2 + 4x + 48}{50}
+ \frac{\sin{(30(x+5))}}{5}.
\end{equation}
The training and validation datasets consist of ground-truth trajectories associated with simulations of 640 input systems generated by sweeping over 10 discrete values of particle mass $m = 1, 2, \ldots, 9, 10$; and 64 uniformly-distributed discrete values of initial position of the particle $x_0 \in [-6.1, 6.1]$. Each simulation produces a trajectory with 400,000 position and velocity values.

\begin{figure}[t]
\centering
\includegraphics[width=0.48\textwidth]{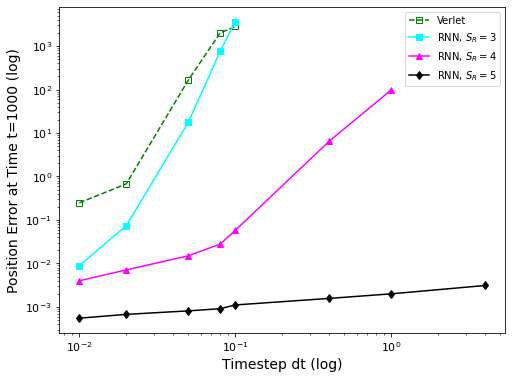}
\caption{
Error $\delta r$ as a function of timestep $dt$ incurred in updating the position of the particle at time $t=1000$ for a 1D system of a particle of mass $m=1$ in a Lennard-Jones potential with initial position $x_0=2.0$. $\delta r$ is evaluated by comparing the predictions to the ground truth results obtained using the Verlet integrator with a small timestep $\Delta$ $=0.001$.
Closed squares, triangles, and diamonds are errors incurred when using RNN operators trained with sequence length $S_{\mathrm{R}}=3,4$ and $5$ respectively.
Open squares correspond to the errors produced by the Verlet integrator. 
}
\label{fig:sequancelength}
\end{figure}

\subsection*{Training RNN Operators with Different Sequence Lengths}

The sequence length $S_{\mathrm{R}}$ is defined as the number of past configurations used by the RNN operator to predict the future configuration. 
Using the training, validation and testing datasets associated with the 1D system of one particle in a Lennard-Jones potential, we did experiments to compare the accuracy of RNN operators designed using $S_{\mathrm{R}} = 3, 4$ and $5$. 
Figure \ref{fig:sequancelength} shows the error $\delta r$ in the prediction of the position of the particle at time $t=1000$ for a system in the test dataset as a function of the timestep $dt$. $\delta r$ is evaluated by comparing the RNN predictions to the ground truth results obtained using the Verlet integrator with a small timestep $\Delta$ $=0.001$.
We find that the RNN operator trained with $S_{\mathrm{R}}=3$ produces errors that rise steeply, spanning over 4 orders of magnitude, as timestep $dt$ is increased from $10\Delta$ to $100\Delta$. The rise in these errors is similar to the increase observed for position errors incurred using the Verlet integrator with increasing $dt$.
The accuracy improves and the errors are comparatively reduced for the RNN operator trained with $S_{\mathrm{R}} = 4$, however $\delta r \gtrsim O(10^{-2})$ for $dt \gtrsim 100\Delta$ and quickly rises to $O(10^{2})$ for $dt = 1000\Delta$. 
In stark contrast, the RNN operator trained with sequence length $S_{\mathrm{R}}=5$ produces errors that stay $O(10^{-3})$ and show a much weaker rise, limited to within an order of magnitude, as $dt$ is increased.


\subsection*{Symplectic property}

\begin{figure*}[t]
\centering
\includegraphics[width=0.95\textwidth]{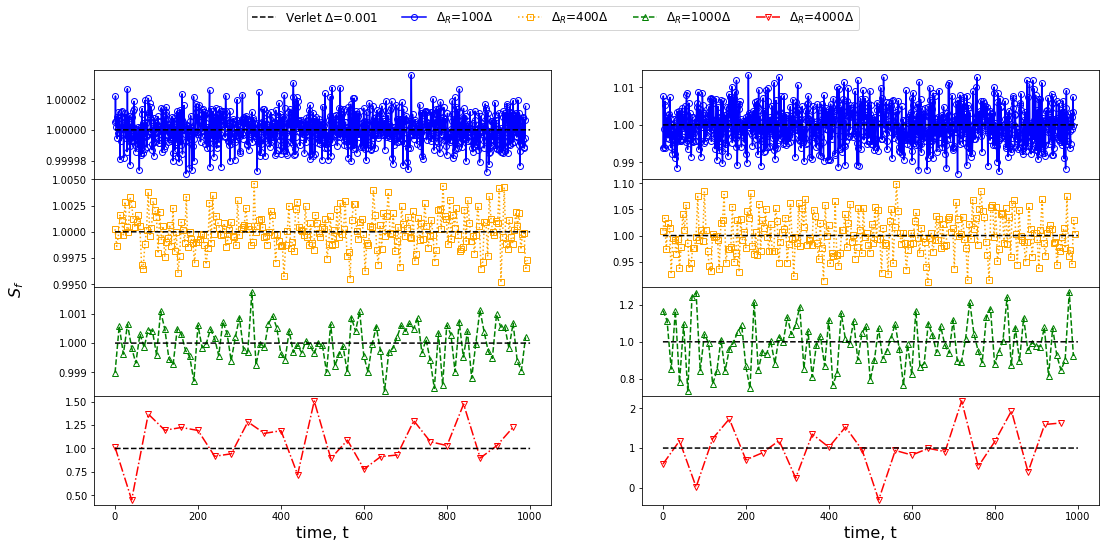}
\caption{Evaluating the preservation of the symplectic property by the RNN operators for 1D systems by computing $S_f$ (defined in Equation \ref{eq:symp} of the main text) from time $t = 0$ to $t = 1000$.
Results are shown for a particle in a simple harmonic potential with parameters $m=10$, $k=1$, $x_0=-10$ (left column), and for a particle in a rugged potential with parameters $m=8$, $x_0=-8$ (right column).
The trajectories are predicted using RNN operators with timestep $\Delta_{\mathrm{R}} = 100\Delta$ (circles), $400\Delta$ (squares), $1000\Delta$ (triangles), and $4000\Delta$ (pentagons). Black lines represent Verlet results with timestep $\Delta = 0.001$. While the Verlet integrator maintains $S_f = 1$ (up to numerical precision), the RNN operators approximately satisfy the symplectic condition ($S_f \approx 1$) up to $1000\Delta$.
Deviations from 1 become much larger for the highest timestep of $4000\Delta$. For a given timestep, fluctuations of $S_f$ near 1 are also larger for the more complex 1D potential (rugged).
}
\label{fig:sympletic_SI}
\end{figure*}

In the main text, we showed that the RNN operators exhibited time-reversal symmetry. In this subsection, we explore numerically whether the trajectories predicted by the RNN operators preserve the symplectic property. We note that these operators are not trained explicitly to preserve the symplectic structure.
For the sake of simplicity, we focus the investigation on 1D systems. In these cases, the symplectic property is obeyed if the trajectories generated using the RNN operators satisfy the equality
\begin{equation}\label{eq:symp}
    JMJ^{T} = M 
\end{equation}
where 
\begin{equation*}
J = \left[\begin{array}{cc}
    \frac{\partial \vec{x}(t)} {\partial \vec{x}(0)} & \frac{\partial \vec{x}(t)} {\partial \vec{p}(0)} \\\\
    \frac{\partial \vec{p}(t)} {\partial \vec{x}(0)} & \frac{\partial \vec{p}(t)} {\partial \vec{p}(0)}
    \end{array} \right]
\end{equation*}
is the Jacobian matrix. Here $\vec{x}(t), \vec{p}(t)$ are the positions and momenta associated with the trajectory of the particles at time $t$, and $\vec{x}(0), \vec{p}(0)$ are the initial positions and momenta at $t = 0$. The matrix $M$ is given by
\begin{equation*}
M = \left[\begin{array}{cc}
\, \vec{0} & \vec{I} \,\, \\
   \, -\vec{I} & \vec{0} \, 
    \end{array} \right]
\end{equation*}
where $0$ and $I$ are $dN\times dN$ dimensional zero and identity matrices, respectively ($d$ is  the spatial dimension and $N$ is the number of particles).
For the case of one particle in 1D, $d = 1$ and $N = 1$, and the matrix $M$ becomes
\begin{equation*}
M = \left[\begin{array}{cc}
   0 & 1   \\
   -1 & 0 
    \end{array} \right].
\end{equation*}
The left hand side of Equation \ref{eq:symp} can be simplified as
\begin{equation*}
  JMJ^{T} = \left[\begin{array}{cc}
    0 &  S_f \\
    -S_f & 0
    \end{array} \right]
\end{equation*}
where $S_f$ is given by:
\begin{equation}\label{eq:symp2}
   S_f = \frac{\partial \vec{x}(t)} {\partial \vec{x}(0)} \frac{\partial \vec{p}(t)} {\partial \vec{p}(0)} -\frac{\partial \vec{x}(t)} {\partial \vec{p}(0)} \frac{\partial \vec{p}(t)} {\partial \vec{x}(0)}. 
\end{equation}
Using the symplectic condition (Equation \ref{eq:symp}), we find
\begin{equation*}
    \left[\begin{array}{cc}
    0 &  S_f \\
    -S_f & 0
    \end{array} \right]
    = \left[\begin{array}{cc}
    0 & 1   \\
    -1 & 0 
    \end{array} \right]
\end{equation*}
Thus, the RNN operators satisfy the symplectic property when 
\begin{equation}
S_f = 1.    
\end{equation}

We computed $S_f$ for the four 1D systems using the Equation \ref{eq:symp2} and the trajectory data predicted by the associated RNN operators with different timestep $\Delta_{\mathrm{R}}$. The ground truth $S_f$ values were obtained using the Verlet integrator with timestep $\Delta = 0.001$. 
We find that for all systems, while the Verlet integrator yields $S_f = 1$ up to the numerical precision, the RNN operators  approximately satisfy the symplectic condition. $S_f$ fluctuates around 1, with the extent of fluctuations becoming stronger with increasing $\Delta_{\mathrm{R}}$ and potential complexity.
Figure \ref{fig:sympletic_SI} shows representative results for a particle in a simple harmonic potential with parameters $m=10$, $k=1$, $x_0=-10$, and for a particle in a rugged potential with parameters $m=8$, $x_0=-8$.
In the case of the particle in a simple harmonic potential, $S_f \approx 1$ for timesteps $\Delta_{\mathrm{R}}=100\Delta, 400\Delta, 1000\Delta$, exhibiting very small fluctuations near 1 (within $\approx 0.1 \%$). However, when the timestep $\Delta_{\mathrm{R}}$ is increased to $4000\Delta$, $S_f$ exhibits greater fluctuations around 1 (deviating by up to $\approx 50\%$). 
Similar trends are observed for the particle in a rugged potential, albeit with greater fluctuations in $S_f$ around $1$ for each $\Delta_{\mathrm{R}}$, which can be attributed to the greater complexity of the rugged potential. 


\subsection*{Energy Profiles for Representative Samples of Few-Particle Systems in Training and Testing Datasets}

\begin{figure}[h]
\centering
\includegraphics[width=0.48\textwidth]{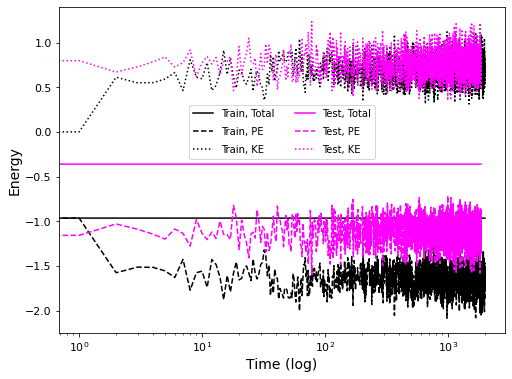}
\caption{
The total, potential, and kinetic energies vs time (log scale) associated with a representative system in the training dataset (black lines) compared with the energy profiles for the test system (magenta lines) of $16$ particles in 3D interacting with Lennard-Jones forces, as described in Figure \ref{fig:manyparticle} of the main text.
For the system in the training dataset, the initial positions are randomly selected, the initial velocities are set to 0, and the Lennard-Jones energy parameter $\epsilon = 1$. For the test system, the initial positions are randomly selected, the initial velocities are sampled from a Boltzmann distribution with a reduced temperature of 1, and $\epsilon = 2$. 
Both sets of energy profiles represent ground truth results obtained using the Verlet integrator with timestep $\Delta=0.001$.
}
\label{fig:manyparticle-Train-energy-profile}
\end{figure}





\begin{thebibliography}{10}
\providecommand{\url}[1]{#1}
\csname url@samestyle\endcsname
\providecommand{\newblock}{\relax}
\providecommand{\bibinfo}[2]{#2}
\providecommand{\BIBentrySTDinterwordspacing}{\spaceskip=0pt\relax}
\providecommand{\BIBentryALTinterwordstretchfactor}{4}
\providecommand{\BIBentryALTinterwordspacing}{\spaceskip=\fontdimen2\font plus
\BIBentryALTinterwordstretchfactor\fontdimen3\font minus
  \fontdimen4\font\relax}
\providecommand{\BIBforeignlanguage}[2]{{%
\expandafter\ifx\csname l@#1\endcsname\relax
\typeout{** WARNING: IEEEtran.bst: No hyphenation pattern has been}%
\typeout{** loaded for the language `#1'. Using the pattern for}%
\typeout{** the default language instead.}%
\else
\language=\csname l@#1\endcsname
\fi
#2}}
\providecommand{\BIBdecl}{\relax}
\BIBdecl

\bibitem{newton}
I.~Newton, \emph{Philosophiae Naturalis Principia Mathematica}.\hskip 1em plus
  0.5em minus 0.4em\relax G. Brookman, 1687.

\bibitem{alder1959studies}
B.~J. Alder and T.~E. Wainwright, ``Studies in molecular dynamics. i. general
  method,'' \emph{The Journal of Chemical Physics}, vol.~31, no.~2, pp.
  459--466, 1959.

\bibitem{frenkel}
D.~Frenkel and B.~Smit, \emph{Understanding Molecular Simulation},
  2nd~ed.\hskip 1em plus 0.5em minus 0.4em\relax Academic Press, 2001.

\bibitem{verlet1967computer}
L.~Verlet, ``Computer" experiments" on classical fluids. i. thermodynamical
  properties of lennard-jones molecules,'' \emph{Physical review}, vol. 159,
  no.~1, p.~98, 1967.

\bibitem{andersen1983rattle}
H.~C. Andersen, ``Rattle: A “velocity” version of the shake algorithm for
  molecular dynamics calculations,'' \emph{Journal of Computational Physics},
  vol.~52, no.~1, pp. 24--34, 1983.

\bibitem{butcher2016numerical}
J.~C. Butcher, \emph{Numerical methods for ordinary differential
  equations}.\hskip 1em plus 0.5em minus 0.4em\relax John Wiley \& Sons, 2016.

\bibitem{wu2016google}
Y.~Wu, M.~Schuster, Z.~Chen, Q.~V. Le, M.~Norouzi, W.~Macherey, M.~Krikun,
  Y.~Cao, Q.~Gao, K.~Macherey \emph{et~al.}, ``Google's neural machine
  translation system: Bridging the gap between human and machine translation,''
  \emph{arXiv preprint arXiv:1609.08144}, 2016.

\bibitem{chong2017deep}
E.~Chong, C.~Han, and F.~C. Park, ``Deep learning networks for stock market
  analysis and prediction: Methodology, data representations, and case
  studies,'' \emph{Expert Systems with Applications}, vol.~83, pp. 187--205,
  2017.

\bibitem{gcfref6}
\BIBentryALTinterwordspacing
X.~Huang, G.~C. Fox, S.~Serebryakov, A.~Mohan, P.~Morkisz, and D.~Dutta,
  ``Benchmarking deep learning for time series: Challenges and directions,'' in
  \emph{2019 {IEEE} International Conference on Big Data (Big Data)}.\hskip 1em
  plus 0.5em minus 0.4em\relax ieeexplore.ieee.org, Dec. 2019, pp. 5679--5682.
  [Online]. Available:
  \url{http://dx.doi.org/10.1109/BigData47090.2019.9005496}
\BIBentrySTDinterwordspacing

\bibitem{ferguson2017machine}
A.~L. Ferguson, ``Machine learning and data science in soft materials
  engineering,'' \emph{Journal of Physics: Condensed Matter}, vol.~30, no.~4,
  p. 043002, 2017.

\bibitem{butler2018machine}
K.~T. Butler, D.~W. Davies, H.~Cartwright, O.~Isayev, and A.~Walsh, ``Machine
  learning for molecular and materials science,'' \emph{Nature}, vol. 559, no.
  7715, p. 547, 2018.

\bibitem{gcfref1}
\BIBentryALTinterwordspacing
{Geoffrey Fox, James A. Glazier, JCS Kadupitiya, Vikram Jadhao, Minje Kim, Judy
  Qiu, James P. Sluka, Endre Somogyi, Madhav Marathe, Abhijin Adiga, Jiangzhuo
  Chen, Oliver Beckstein, and Shantenu Jha}, ``Learning everywhere: Pervasive
  machine learning for effective {High-Performance} computation,'' in
  \emph{{HPDC} Workshop at {IPDPS} 2019}, 2019. [Online]. Available:
  \url{https://arxiv.org/abs/1902.10810}
\BIBentrySTDinterwordspacing

\bibitem{Sharp10943}
\BIBentryALTinterwordspacing
T.~A. Sharp, S.~L. Thomas, E.~D. Cubuk, S.~S. Schoenholz, D.~J. Srolovitz, and
  A.~J. Liu, ``Machine learning determination of atomic dynamics at grain
  boundaries,'' \emph{Proceedings of the National Academy of Sciences}, vol.
  115, no.~43, pp. 10\,943--10\,947, 2018. [Online]. Available:
  \url{https://www.pnas.org/content/115/43/10943}
\BIBentrySTDinterwordspacing

\bibitem{glotzer2017}
\BIBentryALTinterwordspacing
M.~Spellings and S.~C. Glotzer, ``Machine learning for crystal identification
  and discovery,'' \emph{AIChE Journal}, vol.~64, no.~6, pp. 2198--2206, 2018.
  [Online]. Available:
  \url{https://onlinelibrary.wiley.com/doi/full/10.1002/aic.16157}
\BIBentrySTDinterwordspacing

\bibitem{guo2018adaptive}
A.~Z. Guo, E.~Sevgen, H.~Sidky, J.~K. Whitmer, J.~A. Hubbell, and J.~J.
  de~Pablo, ``Adaptive enhanced sampling by force-biasing using neural
  networks,'' \emph{The Journal of chemical physics}, vol. 148, no.~13, p.
  134108, 2018.

\bibitem{botu2015adaptive}
V.~Botu and R.~Ramprasad, ``Adaptive machine learning framework to accelerate
  ab initio molecular dynamics,'' \emph{International Journal of Quantum
  Chemistry}, vol. 115, no.~16, pp. 1074--1083, 2015.

\bibitem{kadupitiya2020ml}
\BIBentryALTinterwordspacing
J.~Kadupitiya, G.~C. Fox, and V.~Jadhao, ``Machine learning for parameter
  auto-tuning in molecular dynamics simulations: Efficient dynamics of ions
  near polarizable nanoparticles,'' \emph{The International Journal of High
  Performance Computing Applications}, 2020. [Online]. Available:
  \url{https://doi.org/10.1177/1094342019899457}
\BIBentrySTDinterwordspacing

\bibitem{long2015machine}
A.~W. Long, J.~Zhang, S.~Granick, and A.~L. Ferguson, ``Machine learning
  assembly landscapes from particle tracking data,'' \emph{Soft Matter},
  vol.~11, no.~41, pp. 8141--8153, 2015.

\bibitem{moradzadeh2019molecular}
A.~Moradzadeh and N.~R. Aluru, ``Molecular dynamics properties without the full
  trajectory: A denoising autoencoder network for properties of simple
  liquids,'' \emph{The journal of physical chemistry letters}, vol.~10, no.~24,
  pp. 7568--7576, 2019.

\bibitem{sun2019deep}
Y.~Sun, R.~F. DeJaco, and J.~I. Siepmann, ``Deep neural network learning of
  complex binary sorption equilibria from molecular simulation data,''
  \emph{Chemical science}, vol.~10, no.~16, pp. 4377--4388, 2019.

\bibitem{aspuru2019}
\BIBentryALTinterwordspacing
F.~Häse, I.~Fdez.~Galván, A.~Aspuru-Guzik, R.~Lindh, and M.~Vacher, ``How
  machine learning can assist the interpretation of ab initio molecular
  dynamics simulations and conceptual understanding of chemistry,'' \emph{Chem.
  Sci.}, vol.~10, pp. 2298--2307, 2019. [Online]. Available:
  \url{http://dx.doi.org/10.1039/C8SC04516J}
\BIBentrySTDinterwordspacing

\bibitem{kadupitiya2019machine}
\BIBentryALTinterwordspacing
J.~Kadupitiya, G.~C. Fox, and V.~Jadhao, ``Machine learning for performance
  enhancement of molecular dynamics simulations,'' in \emph{International
  Conference on Computational Science}, 2019, pp. 116--130. [Online].
  Available:
  \url{https://link.springer.com/chapter/10.1007/978-3-030-22741-8$\_$9}
\BIBentrySTDinterwordspacing

\bibitem{kadupitiya2020machine}
J.~Kadupitiya, F.~Sun, G.~Fox, and V.~Jadhao, ``Machine learning surrogates for
  molecular dynamics simulations of soft materials,'' \emph{Journal of
  Computational Science}, p. 101107, 2020.

\bibitem{wang2019machine}
J.~Wang, S.~Olsson, C.~Wehmeyer, A.~P{\'e}rez, N.~E. Charron, G.~De~Fabritiis,
  F.~No{\'e}, and C.~Clementi, ``Machine learning of coarse-grained molecular
  dynamics force fields,'' \emph{ACS central science}, vol.~5, no.~5, pp.
  755--767, 2019.

\bibitem{raissi2018hidden}
M.~Raissi and G.~E. Karniadakis, ``Hidden physics models: Machine learning of
  nonlinear partial differential equations,'' \emph{Journal of Computational
  Physics}, vol. 357, pp. 125--141, 2018.

\bibitem{long2017pde}
Z.~Long, Y.~Lu, X.~Ma, and B.~Dong, ``Pde-net: Learning pdes from data,''
  \emph{arXiv preprint arXiv:1710.09668}, 2017.

\bibitem{chen2018neural}
T.~Q. Chen, Y.~Rubanova, J.~Bettencourt, and D.~K. Duvenaud, ``Neural ordinary
  differential equations,'' in \emph{Advances in neural information processing
  systems}, 2018, pp. 6571--6583.

\bibitem{endo2018multi}
K.~Endo, K.~Tomobe, and K.~Yasuoka, ``Multi-step time series generator for
  molecular dynamics,'' in \emph{Thirty-Second AAAI Conference on Artificial
  Intelligence}, 2018.

\bibitem{breen2019newton}
P.~G. Breen, C.~N. Foley, T.~Boekholt, and S.~P. Zwart, ``Newton vs the
  machine: solving the chaotic three-body problem using deep neural networks,''
  \emph{arXiv preprint arXiv:1910.07291}, 2019.

\bibitem{chen2019symplectic}
Z.~Chen, J.~Zhang, M.~Arjovsky, and L.~Bottou, ``Symplectic recurrent neural
  networks,'' \emph{arXiv preprint arXiv:1909.13334}, 2019.

\bibitem{shen2017essential}
P.~Shen, X.~Zhang, and Y.~Fang, ``Essential properties of numerical integration
  for time-optimal path-constrained trajectory planning,'' \emph{IEEE Robotics
  and Automation Letters}, vol.~2, no.~2, pp. 888--895, 2017.

\bibitem{raissi2019physics}
M.~Raissi, P.~Perdikaris, and G.~E. Karniadakis, ``Physics-informed neural
  networks: A deep learning framework for solving forward and inverse problems
  involving nonlinear partial differential equations,'' \emph{Journal of
  Computational Physics}, vol. 378, pp. 686--707, 2019.

\bibitem{bar2019learning}
Y.~Bar-Sinai, S.~Hoyer, J.~Hickey, and M.~P. Brenner, ``Learning data-driven
  discretizations for partial differential equations,'' \emph{Proceedings of
  the National Academy of Sciences}, vol. 116, no.~31, pp. 15\,344--15\,349,
  2019.

\bibitem{shen2020deep}
X.~Shen, X.~Cheng, and K.~Liang, ``Deep euler method: solving odes by
  approximating the local truncation error of the euler method,'' \emph{arXiv
  preprint arXiv:2003.09573}, 2020.

\bibitem{raissi2018multistep}
M.~Raissi, P.~Perdikaris, and G.~E. Karniadakis, ``Multistep neural networks
  for data-driven discovery of nonlinear dynamical systems,'' \emph{arXiv
  preprint arXiv:1801.01236}, 2018.

\bibitem{tsai2020learning}
S.-T. Tsai, E.-J. Kuo, and P.~Tiwary, ``Learning molecular dynamics with simple
  language model built upon long short-term memory neural network,''
  \emph{arXiv preprint arXiv:2004.12360}, 2020.

\bibitem{minary2004long}
P.~Minary, M.~Tuckerman, and G.~Martyna, ``Long time molecular dynamics for
  enhanced conformational sampling in biomolecular systems,'' \emph{Physical
  review letters}, vol.~93, no.~15, p. 150201, 2004.

\bibitem{morrone2011efficient}
J.~A. Morrone, T.~E. Markland, M.~Ceriotti, and B.~Berne, ``Efficient multiple
  time scale molecular dynamics: Using colored noise thermostats to stabilize
  resonances,'' \emph{The Journal of chemical physics}, vol. 134, no.~1, p.
  014103, 2011.

\bibitem{leimkuhler2013stochastic}
B.~Leimkuhler, D.~T. Margul, and M.~E. Tuckerman, ``Stochastic, resonance-free
  multiple time-step algorithm for molecular dynamics with very large time
  steps,'' \emph{Molecular Physics}, vol. 111, no. 22-23, pp. 3579--3594, 2013.

\bibitem{chen2018molecular}
P.-Y. Chen and M.~E. Tuckerman, ``Molecular dynamics based enhanced sampling of
  collective variables with very large time steps,'' \emph{The Journal of
  Chemical Physics}, vol. 148, no.~2, p. 024106, 2018.

\bibitem{hochreiter1997long}
S.~Hochreiter and J.~Schmidhuber, ``Long short-term memory,'' \emph{Neural
  computation}, vol.~9, no.~8, pp. 1735--1780, 1997.

\bibitem{chollet2015keras}
F.~Chollet \emph{et~al.}, ``Keras,'' 2015.

\bibitem{buitinck2013api}
L.~Buitinck \emph{et~al.}, ``Api design for machine learning software:
  experiences from the scikit-learn project,'' \emph{arXiv:1309.0238}, 2013.

\bibitem{glorot2010understanding}
X.~Glorot and Y.~Bengio, ``Understanding the difficulty of training deep
  feedforward neural networks,'' 2010, pp. 249--256.

\bibitem{github.RNNMD}
\BIBentryALTinterwordspacing
GitHub, ``Repository rnn-md in softmaterialslab,'' 2021. [Online]. Available:
  \url{https://github.com/softmaterialslab/RNN-MD/}
\BIBentrySTDinterwordspacing

\bibitem{kates2019predicting}
J.~Kates-Harbeck, A.~Svyatkovskiy, and W.~Tang, ``Predicting disruptive
  instabilities in controlled fusion plasmas through deep learning,''
  \emph{Nature}, vol. 568, no. 7753, pp. 526--531, 2019.

\end{thebibliography}

\end{document}